\documentstyle[preprint,aps]{revtex}
\tighten
\begin{document}
\draft
\hyphenation{Nijmegen}

\title{Reply to ``Comment on `Comparison of potential models
       with the \protect\bbox{\lowercase{pp}} scattering data
       below 350 MeV' ''}

\author{V.\ Stoks}
\address{School of Physical Sciences, The Flinders University
         of South Australia, Bedford Park,\\
         South Australia $5042$, Australia}

\author{J.J.\ de Swart}
\address{Institute for Theoretical Physics, University of Nijmegen,
         Nijmegen, The Netherlands}

\date{}
\maketitle

\begin{abstract}
Replying to the above Comment by G.Q.\ Li and R.\ Machleidt, we
point out that the supposed ``flaws'' in our comparison of $N\!N$
potential models with the $pp$ scattering data
[Phys.\ Rev.\ C {\bf 47}, 761 (1993)] are in reality not flaws at all.
\end{abstract}

\narrowtext

In a recent paper~\cite{St93a}, we compared a number of $N\!N$ potential
models with the $pp$ scattering data. The purpose of that paper was to
show that of the numerous potential models that have appeared in the
literature, only a few give a reasonably good description of the $pp$
data. It turns out that only models which have explicitly included the
$pp$ data in their fit give a good description of the $pp$ data.
In general, potential models which have only been fitted to the $np$
data give a very poor description of the $pp$ data.

In a Comment on our paper, Li and Machleidt~\cite{Li93} claim to have
found two flaws in our Comparison of potential models with the $pp$
scattering data.
The first flaw seems to be that, according to Li and Machleidt,
{\it we are not allowed to compare the $I=1$ part of an $N\!N$ potential
to the $pp$ scattering data, when the $^1S_0$ partial wave of that
potential was fitted to the $np$ scattering length}.
However, without this first ``flaw,'' we would never have been able
to come to one of the conclusions of our paper, i.e., that ``only
those potentials that were explicitly fitted to the $pp$ scattering
data give a reasonable description of these data''~\cite{St93a}.

As a next ``flaw'' in our paper, Li and Machleidt claim~\cite{Li93} that
our {\it ``Comparison of $N\!N$ potential models with the $pp$ scattering
data'' has as flaw that it ``compares to the $pp$ data only''.}
To call this a flaw is twisting the logic. In this connection we would
like to point out that our comparison of $N\!N$ potential models with
the $pp$ data was written three years after completion of our
partial-wave analysis of the $pp$ data~\cite{Ber90}. At that moment we
had not yet finished our study~\cite{St93b} of the $np$ data. This
means that it was at that moment too early to expect from the Nijmegen
group a comparison of $N\!N$ potentials with the $np$ data.

We will use the remainder of this Reply to elaborate on these points
and in turn address a few flaws in the Comment.

The Comment states as the most important rule, that ``a $pp$
potential must only be confronted with the $pp$ data, while an $np$
potential must only be confronted with the $np$ data''~\cite{Li93},
where their main criterion for the denomination $pp$ or $np$ is
whether the $^1S_0$ partial wave of the potential has been fitted to
the $pp$ or $np$ scattering length, respectively.
This rule sounds reasonable and noble, but is actually so often
violated that it is presumptious to call it a rule.
In most practical situations (e.g., in three-body, nuclear matter, and
$pp$ bremsstrahlung calculations), one usually starts with some nuclear
potential model to represent the $N\!N$ interaction (for $pp$, $np$,
and $nn$), disregarding the details whether its parameters were
fitted to the $pp$ and/or $np$ data. In our paper~\cite{St93a}, we
explicitly demonstrated that this is incorrect.

Moreover, we feel that with this first flaw we are in good company,
because we have it in common with Machleidt {\it et al.} in
Ref.~\cite{Mac87}. In this paper on the Bonn $N\!N$ potential, it is
stated that ``one measures $N\!N$ observables (and not phase shifts),''
and that therefore ``the real test of the quality of an $N\!N$
interaction is the comparison with these data.
For their model this is done in their Fig.~16''.
Whereas we do a {\it quantitative\/} comparison~\cite{St93a} to the
data, they merely do a {\it qualitative\/} comparison in their Fig.~16,
from which they then claim that it can be ``seen'' that an ``excellent
description of the data is achieved''~\cite{Mac87}. In this Fig.~16,
these authors compare 10 times with the $np$ data and 6 times with the
$pp$ data. A curious double standard:
When Machleidt {\it et al.}~\cite{Mac87} compare with the $pp$ data,
it is allowed, whereas when we compare with the $pp$ data, it is called
a flaw.

With respect to this Ref.~\cite{Mac87}, we would also like to point
out that they write about the ``$N\!N$ potential, $N\!N$ interaction,
$N\!N$ observables,'' and {\it not\/} about the ``$np$ potential,
$np$ interaction, $np$ observables.''

In spite of their own rule, Li and Machleidt themselves write that it
is valid to confront an $np$ potential with the $pp$ data. The only
changes one has to make are ``to include the Coulomb force and to
readjust the $^1S_0$ scattering length to its $pp$ value''~\cite{Li93}.
They suggest this can be achieved by a relatively small change in one
of the model parameters. As an example, they state that ``a change of
the $\sigma$ coupling constant by as little as 1\% changes the
$\chi^2/N_{\rm data}$ from 641 to 2''~\cite{Li93}. They have in mind
here the full Bonn potential~\cite{Mac87}. For the reader not familiar
with the field, we would like to show how misleading these arguments
of Li and Machleidt are.

First of all, the authors of the Comment fail to mention that we
already showed the above example to be inaccurate in the general case.
In our paper~\cite{St93a} we show that when we replace the $^1S_0$
phase shift of the Argonne potential~\cite{Wir84} (fitted to the $np$
scattering length) by the $^1S_0$ phase shift of the Nijmegen
multi-energy analysis~\cite{Ber90} (which roughly corresponds to having
a model with ``perfect'' $^1S_0$ phase shifts), then the quality of
the model improves considerably.
However, ``the resulting $\chi^2/N_{\rm data}\approx 4$ is still rather
large,'' demonstrating that ``the other phase shifts are not too good
either''~\cite{St93a}. As is shown in Table~\ref{table1}, the situation
for the various coordinate-space Bonn potentials is worse. In this Table
we show the effect of replacing the $^1S_0$ phase shifts of the potential
models by the $^1S_0$ phase shift of the latest Nijmegen partial-wave
analysis~\cite{St93b}. Clearly, the quality of these coordinate-space
Bonn potentials is rather poor.

Secondly, in our potential comparison~\cite{St93a} we did not include
the full Bonn $np$ potential, but only the coordinate-space version
OBEPR. The $\chi^{2}/N_{\rm data}$ of this Bonn potential starts at
492 (and not at 641) and after improving the $^1S_0$ phase shift by
refitting, the $\chi^{2}/N_{\rm data}$ drops to 7.1 (and not 2). The
$\chi^{2}/N_{\rm data}$ of this Bonn coordinate-space $np$ potential
OBEPR of 1987 is worse than that of the 25 years older Hamada-Johnston
potential~\cite{Ham62} of 1962. This in spite of the fact that the
parameters of OBEPR were fitted~\cite{Mac87} to arrive at a
``realistic'' description of the $N\!N$ scattering data.

At this point we like to mention that we disagree with Li and
Machleidt that the $\chi^2$ of the coordinate-space Bonn potentials
is of no interest. In a footnote they write that these potentials
were merely constructed ``to point out the deficiencies of such simple
models''~\cite{Li93}. But when that is true, why at all bother to
present a coordinate-space version of the one-boson-exchange version
of the Bonn potential and then two years later again present~\cite{Mac89}
coordinate-space versions of both the Bonn A and Bonn B potentials?
Moreover, as can be seen in Table II of our comparison~\cite{St93a},
there exist other similar coordinate-space potential models which do
give a reasonable description of the scattering data.
We think that the users of the coordinate-space Bonn potentials should
be warned, that these potentials do {\it not\/} give a reasonable fit
to the $pp$ scattering data and, therefore, these potentials should not
be used to calculate, e.g., $pp$ bremsstrahlung or the binding energy
of the triton.

In their footnote 10, Li and Machleidt~\cite{Li93} find even fault with
the energy range used by us. When these authors~\cite{Mac93} fit the
Bonn potentials to the $pp$ data they use the Nijmegen partial-wave
analysis. Then in their Comment these authors proceed to point out, as
if we did not already know, ``that the $pp$ data carry a very small error
at low energies''~\cite{Li93} and, when included, add disproportionally
to the total $\chi^{2}$. Of course, we are well aware of these facts.
This was precisely the reason why in our comparison~\cite{St93a} we
{\it also\/} explicitly presented the results for the case where we
removed these low-energy data.
Their lack of knowledge about the $N\!N$ data is shown when, without
any supporting evidence, they state that ``the real potential concept
is strictly speaking wrong above 280--290 MeV. It may be o.k.\ to stretch
the limit by a few MeV up to 300 MeV, but not beyond that''~\cite{Li93}.
This is absolutely incorrect. Because of bremsstrahlung, the real
potential concept is strictly speaking wrong at any energy $E>0$.
When we include in our partial-wave analysis
of the 0--350 MeV $pp$ data the inelasticities as determined in a 0--500
MeV $pp$ partial-wave analysis~\cite{Kok93}, then we see a drop in
$\chi^{2}/N_{\rm data}$ of about 0.0006, which is truly
negligible. This shows that the real potential concept is
valid to {\it at least\/} 350 MeV and definitely a fair distance beyond that
energy as well. Another way to see this is to look at Table II of
Ref.~\cite{St93a}. There it can be seen that most potentials fit the
$pp$ data in the interval 290--350 MeV {\it better\/} than over the
whole energy range 2--300 MeV. A notable exception, however, is the full Bonn
$pp$ potential~\cite{Hai89}, which in this 290--350 MeV region has a
$\chi^{2}/N_{\rm data}$ which is {\it worse} than average.
When a potential does not fit well the data in this energy interval, then
it is the potential which is at fault, and not the potential concept.

Let us conclude with some additional remarks on why we only made a
comparison with the $pp$ data.
It is not surprising that the more
recent potentials (e.g., the Argonne~\cite{Wir84} and Bonn~\cite{Mac87}
potentials) were fitted only to the $np$ data. It is much easier to
fit a potential model to the $np$ scattering data than it is to the
$pp$ data, because in that case it is not necessary to go through all
the difficulties of making all kinds of electromagnetic corrections.
And as already stressed before, it is often assumed that with only
a minor adjustment in one of the parameters of the model, that model
will also give a good description of the $pp$ data.
By explicitly confronting some of these $np$ potentials with the $pp$
data, we showed~\cite{St93a} that this assumption is not always valid.
Even readjusting the $^1S_0$ phase shift is often not good enough.
So it is definitely {\it not true} to state that ``a potential that
fits the $np$ phase shifts well, will automatically fit the $pp$ phase
shifts well'', and that ``there is no need to bring the $pp$ data
into play''~\cite{Mac89}.

Moreover, one has to bear in mind that the older potential models were
fitted using old databases. For example, the Nijm78 and Paris80 potentials
[12,13] 
were fitted to the 1969 Livermore
database~\cite{Mac69}. At that time the $pp$ data were already accurate
enough to pin down the $I=1$ partial waves. So well even that at the
present time these  Nijm78 and Paris80 potentials still give a good
description of the $pp$ data, including the more accurate recent data.
On the other hand, the quality of the $np$ data was very poor at that
time. In the last two decades the quality of the $np$ data has improved
enormously, especially in providing more accurate bounds on the $I=0$
partial waves. Hence, it is not so surprising that the old Nijm78 and
Paris80  potentials give a rather poor description of the present $np$ data,
whereas the full Bonn87 potential is claimed to fit these recent $np$ data
much better. It should be noted, however, that by simply refitting the
parameters, the Nijm78 and Paris80 potentials are likely to describe
the $np$ data equally well as the $pp$ data. As a matter of fact, an
updated~\cite{St93c} version, Nijm92, of the Nijmegen potential describes all
$N\!N$ scattering data ($pp$ and $np$) with $\chi^2/N_{\rm data}=1.9$.

Therefore, it is incorrect to compare the $I=0$ partial waves of the
Nijm78, Paris80, and Bonn87 potentials, and then (because of the large
differences) come to the conclusion that~\cite{Mac93} ``The more
seriously and consistently meson theory is pursued, the better the
results.'' The $\chi^{2}$ is the product of the number and kind of
free parameters in the model and the effort one wants to invest in
fitting the data. The increase in computing power over the last decade,
together with the ease of access to scattering data, to partial wave
analyses, and their improved
accuracy, has made it possible to construct easily much better
potentials~\cite{St93c} than the Nijm78, Paris80, and Bonn87 potentials.
This makes a farce of the incorrect claim by Machleidt et al.\
that ``their better results indicate, that
they apply meson theory more seriously and consistently''~\cite{Mac93}.

In summary, our earlier paper~\cite{St93a} gives a fair comparison of
the quality of a number of $N\!N$ potential models with respect to the
$pp$ scattering data. It clearly demonstrates the fact that potentials
which have only been fitted to the $np$ data do not automatically give
a good description of the $pp$ data, even after adjusting the $^1S_0$
partial wave. In fact, most potential models do not.
Therefore, our paper~\cite{St93a} perfectly serves its purpose:
namely, to point out this misconception, and to warn other physicists to
administer some care in using these models in practical calculations.

\narrowtext
\begin{table}
\caption{$\chi^2/N_{\rm data}$ results in the 2--350 MeV range for the
         coordinate-space OBEPR~\protect\cite{Mac87}, Bonn A and
         Bonn B~\protect\cite{Mac89} potentials.
         In the second line the $^1S_0$ of the potential is replaced
         by the $^1S_0$ of the Nijmegen analysis~\protect\cite{St93b}.}
\begin{tabular}{crrr}
  $^1S_0$     &  OBEPR &  Bonn A &  Bonn B  \\
\tableline
  potential   &  13.4  &   10.4  &    9.2   \\
  analysis    &   8.1  &    6.8  &    5.5   \\
\end{tabular}
\label{table1}
\end{table}


\begin{references}
\bibitem{St93a} V.\ Stoks and J.J.\ de Swart,
         Phys.\ Rev.\ C {\bf 47}, 761 (1993).
\bibitem{Li93}  G.Q.\ Li and R.\ Machleidt,
         submitted to Phys.\ Rev.\ C.
\bibitem{Ber90} J.R.\ Bergervoet, P.C.\ van Campen, R.A.M.\ Klomp,
         J.-L.\ de Kok, T.A.\ Rijken, V.G.J.\ Stoks, and J.J.\ de Swart,
         Phys.\ Rev.\ C {\bf 41}, 1435 (1990).
\bibitem{St93b} V.G.J.\ Stoks, R.A.M.\ Klomp, M.C.M.\ Rentmeester,
         and J.J.\ de Swart,
         Phys.\ Rev.\ C, to appear.
\bibitem{Mac87} R.\ Machleidt, K.\ Holinde, and Ch.\ Elster,
         Phys.\ Rep.\ {\bf 149}, 1 (1987).
\bibitem{Wir84} R.B.\ Wiringa, R.A.\ Smith, and T.L.\ Ainsworth,
         Phys.\ Rev.\ C {\bf 29}, 1207 (1984).
\bibitem{Ham62} T.\ Hamada and I.D.\ Johnston,
         Nucl.\ Phys.\ {\bf 34}, 382 (1962).
\bibitem{Mac89} R.\ Machleidt,
         Adv.\ Nucl.\ Phys.\ {\bf 19}, 189 (1989).
\bibitem{Mac93} R.\ Machleidt and G.Q.\ Li,
         {\it Nucleon-Nucleon Potential in Comparison: Physics or
         Polemics?}, talk presented at ``Realistic Nuclear
         Structure'', a conference to mark the 60th birthday of
         T.T.S.\ Kuo, May 1992, Stony Brook, New York; unpublished.
\bibitem{Kok93} J.-L.\ de Kok, private communication.
\bibitem{Hai89} J.\ Haidenbauer and K.\ Holinde,
         Phys.\ Rev.\ C {\bf 40}, 2465 (1989).
\bibitem{Nag78} M.M.\ Nagels, T.A.\ Rijken, and J.J.\ de Swart,
         Phys.\ Rev.\ D {\bf 17}, 786 (1978).
\bibitem{Lac80} M.\ Lacombe, B.\ Loiseau, J.M.\ Richard, R.\ Vinh Mau,
         J.\ C\^ot\'e, P.\ Pir\`es, and R.\ de Tourreil,
         Phys.\ Rev.\ C {\bf 21}, 861 (1980).
\bibitem{Mac69} M.H.\ MacGregor, R.A.\ Arndt, and R.M.\ Wright,
         Phys.\ Rev.\ {\bf 182}, 1714 (1969).
\bibitem{St93c} V.G.J.\ Stoks, R.A.M.\ Klomp, and J.J.\ de Swart,
         in preparation.
\end{references}
\end{document}